\documentclass[reprint,footinbib,amsmath,amssymb,aps,pre,floatfix,superscriptaddress,a4paper]{revtex4-2}
\usepackage{graphicx}
\usepackage{dcolumn}
\usepackage{bm}
\usepackage[colorlinks=true,allcolors = blue]{hyperref}
\usepackage{newtxtext,newtxmath}
\usepackage{siunitx}
\usepackage{microtype}
\usepackage{selnolig}
\usepackage{bbold} 
\usepackage[dvipsnames]{xcolor}

\newcommand{\uav}[1]{\langle{#1}\rangle}
\newcommand{\biguav}[1]{\Bigr\langle{#1}\Bigl\rangle}
\newcommand{\load}{N}
\newcommand{\drag}{F}

\newcommand{\rmd}[1]{\mathrm{d}{#1}} 
\newcommand{\order}[1]{\mathcal{O}({#1})}
\newcommand{\leibnizd}[1]{\mathrm{d}{#1}}
\newcommand{\fubar}[1]{\frac{#1}{\sqrt{z}}\tan^{-1}\!\frac{1}{\sqrt{z}}}

\newcommand{\partFig}[2]{Fig.~\hyperref[#1]{\ref*{#1}#2}}
\newcommand{\ecfp}{Edinburgh Complex Fluids Partnership and School of Physics and Astronomy, The~University~of~Edinburgh, James Clerk Maxwell Building, Peter Guthrie Tait Road, Edinburgh EH9 3FD, United Kingdom}

\newcommand{\Eqref}[1]{Eq.~\eqref{#1}}
\newcommand{\Eqsref}[1]{Eqs.~\eqref{#1}}
\newcommand{\Figref}[1]{Fig.~\ref{#1}}

\newcommand{\Secref}[1]{\S~\ref{#1}}
\newcommand{\Refcite}[1]{Ref.~\onlinecite{#1}}
\newcommand{\partref}[2]{\hyperref[#1]{#2}}
\newcommand{\partFigref}[2]{Fig.~\hyperref[#1]{\ref*{#1}#2}}

\newcommand{\latin}[1]{{\itshape #1}}

\newcommand{\etal}{\latin{et al.}}
\newcommand{\eg}{\latin{e.g.}}
\newcommand{\ie}{\latin{i.e.}}

\newcommand{\viz}{\latin{viz.}}

\begin{document}
\title{Gap-Dependent Hydrodynamic Lubrication in Conformal Contacts}
\author{James A. Richards}
\email{james.a.richards@ed.ac.uk}
\affiliation{\ecfp}
\author{Patrick B. Warren}
\email{patrick.warren@stfc.ac.uk}
\affiliation{The Hartree Centre, STFC Daresbury Laboratory, Warrington WA4 4AD, United Kingdom}
\author{Wilson C. K. Poon}
\email{w.poon@ed.ac.uk}
\affiliation{\ecfp}
%
%
\begin{abstract}
We show that the hydrodynamic lubrication of contacting conformal surfaces with a typical texture height gives rise to a universal behaviour in the Stribeck curve in which the friction coefficient shows an anomalous power-law dependence on the Sommerfeld number, $\mu \sim S^{2/3}$. When the gap height drops below the `texture length scale', deviations from $S^{2/3}$ occur, which may resemble the onset of elasto-hydrodynamic and boundary lubrication. Within this framework, we analyse literature data for oral processing and find $S^{2/3}$ scaling with deviations consistent with measured lengthscales.
\end{abstract}
\maketitle


\section{Introduction\label{sec:pre:calcNon}}

\begin{figure}[t]
    \centering
    \includegraphics{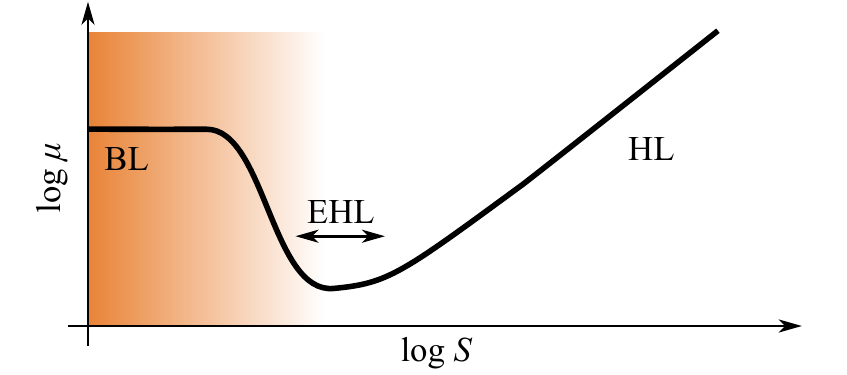}
    \caption{Schematic Stribeck curve, friction coefficient ($\mu = \drag/\load$) as a function of Sommerfeld number ($S=\eta U \load/R$, for sliding speed $U$, lubricant viscosity $\eta$ and radius of curvature of lubrication geometry $R$). Regimes of $\mu$ with decreasing $S$: hydrodynamic lubrication (HL), elasto-hydrodynamic lubrication (EHL) at minimum, and constant boundary lubrication (BL, shading).}
    \label{fig:stribeck}
\end{figure}

The importance of lubricated contacts between sliding surfaces cannot be overstated~\cite{hamrock2004fundamentals,Hirani2016}. Such contacts are often characterised by a `pin-on-disc test', which is analysed in terms of a hemisphere of radius $R$ trapping a lubricant of viscosity $\eta$ sliding on a flat surface. The friction coefficient $\mu$, the ratio of the imposed normal force $\load$ on the pin to the tangential drag force $\drag$ on the disc, $\mu = \frac{\drag}{\load}$, is measured as $\load$ and the relative sliding speed $U$ vary, giving $\mu$ as a function of the Sommerfeld number $S = \eta U R/N$~\cite{gumbel1914problem,hersey1914laws,stribeck1902abc}, a Stribeck curve.

The typical Stribeck curve shows three regimes, \Figref{fig:stribeck}. At high $S$, $N$ is purely hydrodynamic in origin~\cite{reynolds1886theory}. In this hydrodynamic lubrication (HL) regime, $\mu$ decreases as $\load$ increases, reaching a minimum before upturning due to the onset of elastohydrodynamic lubrication (EHL), where hydrodynamic stresses begin to deform surface asperities, which however are not yet touching. Touching signals the onset of the boundary lubrication (BL), where $\mu$ becomes nearly constant as $S\to 0$.

\begin{figure}[t]
    \centering
    \includegraphics{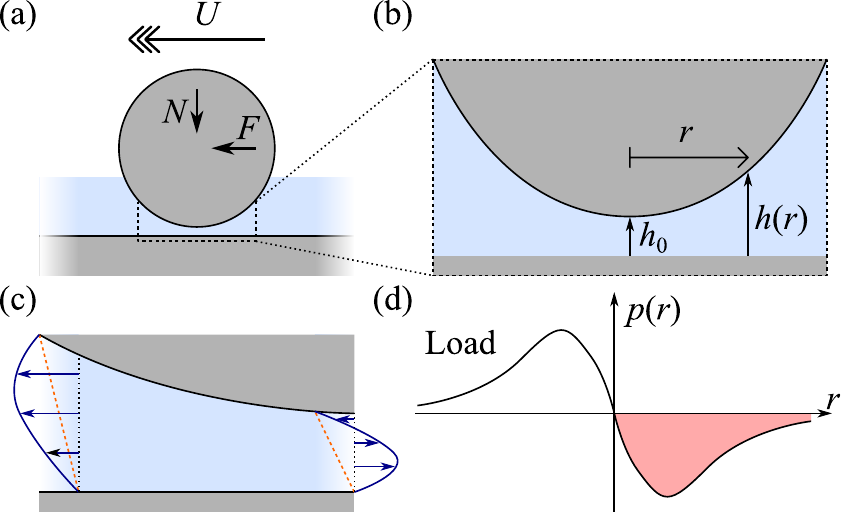}
    \caption{Non-conformal lubrication. (a)~Lubrication geometry, sphere or cylinder with radius $R$, loading conditions with normal ($\load$) and tangential load ($\drag$) applied to upper surface [dark (grey) shading], with opposing forces from fluid [light (blue) shading] due to relative sliding speed ($U$) between upper and lower surfaces. (b)~Narrow-gap region with gap $h(r)$ at a distance $r \ll R$ from point of closest approach with gap $h_0$. Curved surface in narrow-gap region approximated by parabola, $h(r) = h_0 + r^2/2R$. (c)~Schematic of resulting fluid flow. Couette flow [(orange) dashed lines] with changing gap results in compensating Poiseuille flow to give changing net fluid flow [(blue) solid lines and arrows]. (d)~Resulting Reynolds lubrication pressure distribution $p(r)$ schematic. Symmetry implies equal and opposite pressures for $r \to -r$, negative pressure [(red) shading] neglected due to cavitation, the half-Sommerfeld boundary condition.}
    \label{fig:pre:cylinder}
\end{figure}

Hydrodynamic lubrication originates from volume conservation in the convergent thin-film flow of an incompressible lubricant~\cite{rayleigh1918notes}. Consider a ball-on-flat geometry, Fig.~\ref{fig:pre:cylinder}a-b, with a plane Couette flux $Uh/2$ (volume per unit width, here and throughout) in the convergent gap due to the ball sliding along $x$ relative to the plane at velocity $-U$. This varies with the gap, \partFigref{fig:pre:cylinder}{c} (dashed red), and does not conserve volume. The imbalance is corrected by a gap-dependent pressure-driven in-plane Poisueille flux, $-\frac{h^3}{12\eta}\nabla p$, \partFigref{fig:pre:cylinder}{c} (black). The associated pressure field $p$ provides the normal force to keep the surfaces apart. 

Mathematically, balancing the gap-dependent Couette ($\mathbf{J}_c$) and Poisseuille ($\mathbf{J}_p$) fluxes translates to $\nabla \cdot (\mathbf{J}_c + \mathbf{J}_p) = 0$. Noting that $\mathbf{J}_c$ is exclusively in the $x$ direction while $\mathbf{J}_p$ can in general lie in the $(x,y)$ plane, we therefore require
\begin{equation}
  \frac{\partial}{\partial x}\Bigl(\frac{Uh}{2}\Bigr)- \left[\frac{\partial}{\partial x}\Bigl(\frac{h^3}{12\eta}\frac{\partial p}{\partial x}\Bigr)
  + \frac{\partial}{\partial y}\Bigl(\frac{h^3}{12\eta}\frac{\partial p}{\partial y}\Bigr) \right] =0\,. \label{eq:pre:pressure}
\end{equation}
In this Reynolds lubrication equation~\cite{reynolds1886theory}, the pressure $p(x,y)$ is independent of the vertical coordinate because the gap $h(x,y)$ is everywhere small compared to the flow's length and width. Over the bearing area, this pressure supports $\load$, while $\drag$ has Couette and Poiseuille components:
\begin{equation}
  \load=\int\!\!\rmd x \,\rmd y\,p\,,\quad
  \drag=\int\!\!\rmd x \,\rmd y\left[\frac{\eta U}{h}+\frac{h}{2}\left(\frac{\partial p}{\partial x}+\frac{\partial p}{\partial y}\right)\right]. \label{eq:pre:loaddrag}
\end{equation} 

Analytic solutions to \Eqref{eq:pre:pressure} exist only for a few cases. For a ball (radius $R$) on a flat surface with minimum gap $h_0\ll R$, \partFigref{fig:pre:cylinder}{a}, Kapitza gave $h_0 = (72\pi^2/25) RS^2 \approx 28.4RS^2$~\cite{kapitza1955hydrodynamic} while Hamrock has $h_0 \approx 34.8 RS^2$~\cite{hamrock2004fundamentals}, reflecting different approximations. They also calculated $p(x,y)$. Similar results for other canonical geometries are presented in textbooks~\cite{Hirani2016,hamrock2004fundamentals,Faber1995,Guyon2015}. Curiously, however, there are few calculations of and experimental data for $\mu(S)$ in the HL regime. So, while HL is widely thought to be understood, much of the theory remains untested by direct experiments. 

In an accompanying paper~\cite{PRL}, we address this knowledge gap for conformal contacts, \ie\ surfaces with commensurate curvature. We outline scaling arguments predicting an unexpected $\mu \sim S^{2/3}$ behaviour at large $S$ in the HL regime, and deviations below some critical $S^\star$. Experiments confirm these predictions. Here, we justify our scaling arguments and present full calculations, paying particular attention to deviations from scaling around $S^\star$, which may resemble the transition to EHL. Below, after recalling the scaling theory for the ball-on-flat geometry~\cite{warren2016sliding}, we report detailed scaling analyses for conformal contacts to reveals the physics underlying the  $S^{2/3}$ scaling and deviations from it at low $S$. Solving \Eqref{eq:pre:pressure} by quadrature follows to explore these deviations in depth. We end by applying our results to some recently-published data~\cite{soltanahmadi_insights_2023} to demonstrate the practical utility of our analysis.

\section{Scaling theory of non-conformal contacts} \label{sec:pre:scaleNonconform}

The two surfaces in a non-conformal contact have different curvatures. Kapitza~\cite{kapitza1955hydrodynamic} solved the sphere on flat, Fig.~\ref{fig:pre:cylinder}a, analytically. The physics emerges from a scaling argument~\cite{warren2016sliding}.

First, expand the convergent gap around its narrowest point, $h(r) \approx h_0 + r^2/2R$, \partFigref{fig:pre:cylinder}{b}. In a region of radius $r_0 \sim \sqrt{h_0 R}$, the gap remains narrow, and we expect from \Eqref{eq:pre:pressure} that
\begin{equation}
    \frac{Uh_0}{r_0} \sim \frac{1}{r_0} \times \frac{h_0^3}{\eta} \frac{p}{r_0} \quad\Rightarrow\quad p \sim \frac{\eta U r_0}{h_0^2}\,. \label{eq:pre:ballScale}
\end{equation}
This supports load $N \sim p r_0^2 \sim \eta U R^{3/2}/h_0^{1/2}$, or $h_0/R \sim (\eta U R/\load)^2 \sim S^2$, confirming Kapitza~\cite{kapitza1955hydrodynamic}.
Symmetry predicts equal but opposite pressure in the diverging gap. If this negative pressure is significantly below atmospheric and the lubricant cavitates, then the `half Sommerfeld boundary condition' of $p = 0$ applies, giving a net normal force. 

The narrow-gap shear rate $\dot\gamma \sim U/h_0$ gives a shear stress $\sigma \sim \eta U /h_0$ and lateral force $\drag \sim \sigma r_0^2 \sim \eta U R$, so that
\begin{equation}
    \mu = \frac{\drag}{\load} \sim \frac{\eta U R}{\load} \sim S\,. 
\end{equation}
Such linear scaling is often assumed in sketches of the Stribeck curve in the HL regime (\eg,  Fig.~2 in \Refcite{woydt2010history}).

For a long cylinder, $x_0 \sim \sqrt{h_0R}$ in the narrow gap region. If its length $L$ satisfies $L \gg x_0$, then lubricant flow is exclusively along the direction of motion rather than through side leakage. The pressure has the same form as the ball-on-flat case, but now acts over area $\sim x_0L$, not $r_0^2$, which support a load $\load \sim x_0L p_0 \sim x_0^2 \eta U L/h_0^2 \sim \eta U L(R/h_0)$, with associated drag $\drag \sim \sigma x_0 L \sim \eta U L \sqrt{R/h_0}$. Now we predict
\begin{equation}
  S 
  \equiv \frac{\eta U L}{\load} \sim \frac{h_0}{R}\,,\quad \mu \sim \sqrt{\frac{h_0}{R}} \quad\Rightarrow\quad \mu \sim S^{1/2}\,.
\end{equation}

Apparently, the HL of non-conformal contacts gives $\mu \sim S^\alpha$ with $\alpha$ set by the spatial dimension. However, few published Stribeck curves show either of these scalings. 

\section{Scaling theory of conformal contacts\label{sec:pre:calcConf}}

\begin{figure}[t]
    \centering
    \includegraphics[width=0.8\columnwidth]{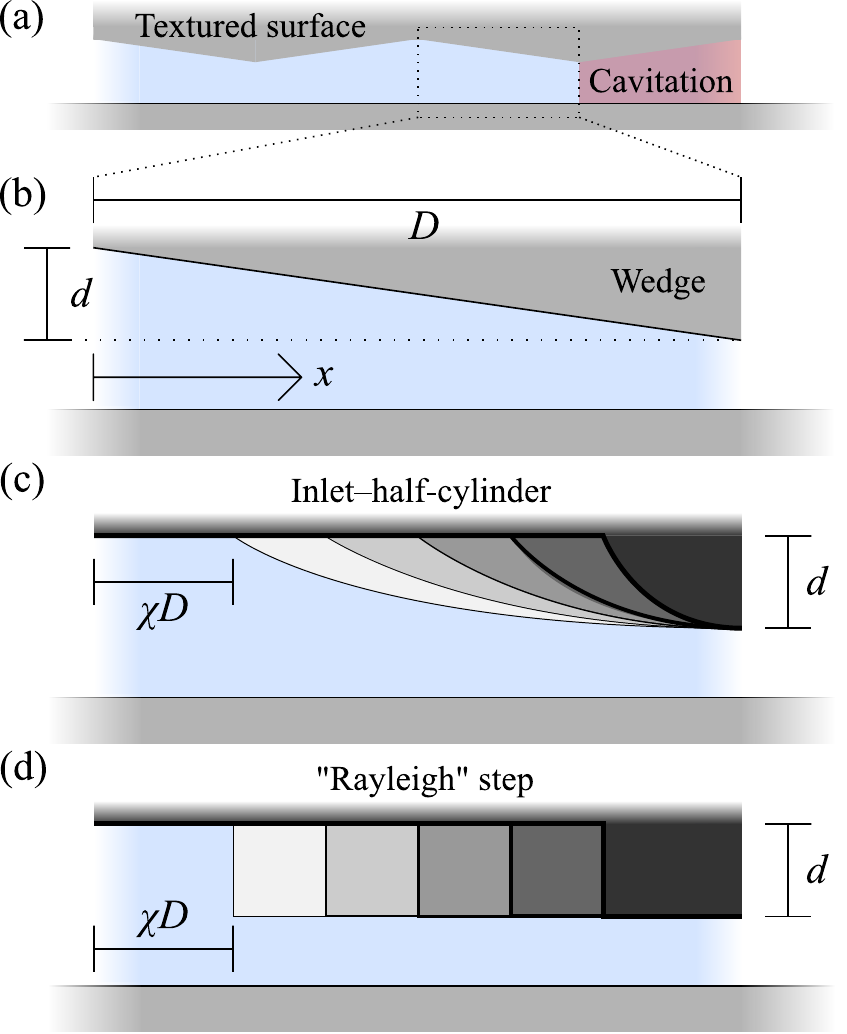}
    \caption{(a) A textured surface of back-to-back wedge bearings. (b)~Wedge profile specified by length $D$ and `step height' $d$ as decrease in gap ($h_0+d$ to $h_0$) from $x=0$ to $x=D$. Dimensionless gap profile $\xi = z +1 - u$ for scaled co-ordinates, $z = h_0/d$ and $u = x/D$, \Eqref{eq:pre:wedgeProfile}. (c)~Inlet--half-cylinder profile. Cylinder approximated by parabolic form from $u = \chi$ to 1, \Eqref{eq:pre:inletProfile}, inlet region, $h_0 +d$ at $u=0$ to $\chi$ (or $x = \chi D$). Increasing $\chi$, and larger inlet, shown from light to dark shading and with fine to bold lines. (d)~`Rayleigh' step gap profile with equivalent bearing dimensions, \Eqref{eq:pre:stepProfile}.}
    \label{fig:pre:steps}
\end{figure}

The ball or cylinder on plane is characterized by a single length scale, \viz, the sphere or cylinder radius, $R$. A contact between two equal-curvature surfaces apparently has no intrinsic length scale. However, the surfaces in a real conformal contact will show a degree of mesoscale non-flatness with characteristic length scale $d \gg$ typical asperities. They are therefore `textured surfaces'~\cite{gropper2016hydrodynamic} in which the gap height varies from some minimum $h_0$ to $h_0 + d$ over the area of contact, \partFigref{fig:pre:steps}{a}. We restrict ourselves to surface undulations along the sliding direction of the conformal contact. Assuming the  half-Sommerfeld boundary condition, we consider only the converging gap regions, each of which is then a slider bearing of length $D \gg d$ and width $L \gg d$.

\begin{figure}[t]
    \centering
    \includegraphics[width=\columnwidth]{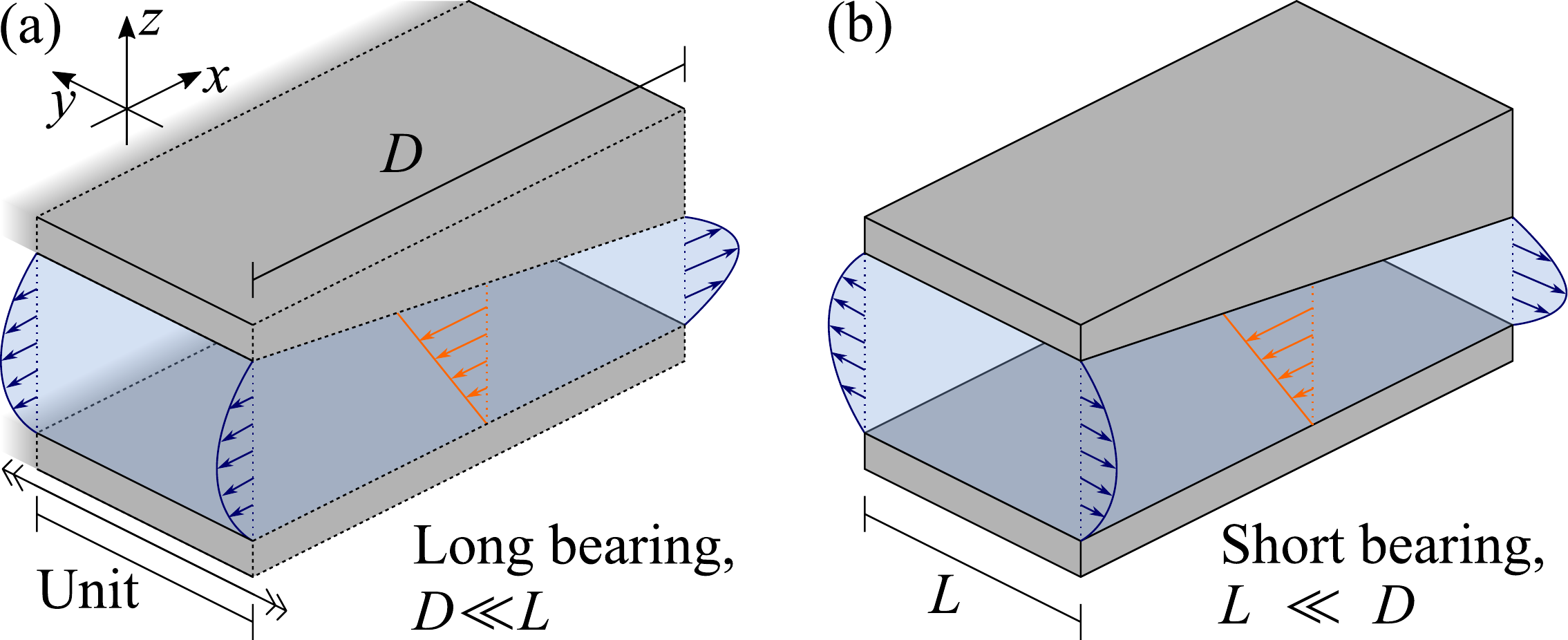}
    \caption{Fluid flow in slider bearings. (a)~Long-bearing limit with length much less than width ($D\ll L$) resulting in compensating Poiseuille flow [dark(blue) arrows] in same $x$ direction as Couette shear flow [light (orange) arrows)]. (b)~Corresponding short-bearing limit with length much greater than width ($D \gg L$) leading to ``side-leakage'' in $y$ direction perpendicular to Couette flow.}
    \label{fig:pre:types}
\end{figure}

For the three slider bearings in \Figref{fig:pre:steps}, a wedge, a `Rayleigh step', and a half-cylinder inlet, there is a characteristic `step height', $d$. Two limits exist, \Figref{fig:pre:types}, the `long bearing', $L \gg D$ (\partFigref{fig:pre:types}{a}), or the the `short bearing', $L \ll D$ (\partFigref{fig:pre:types}{b}), where volume conservation is ensured by side-leakage~\cite{dubois1953analytical}. In each case, there is a large-gap regime ($h_0 \gg d$) and a small-gap regime ($h_0 \ll d$), with cross-over at $h_0/d \approx 1$.

Full calculations, Sections~\ref{sec:pre:exact} and \ref{sec:pre:alpha}, predict the Stribeck curves shown in \Figref{fig:pre:stribeck}. To highlight the physics, we first give scaling analyses for the universal power-law behaviour at high $S$ and for the details-dependent limit $S\to 0$.

\subsection{Large-gap limit, $h_0 \gg d$}

For a long slider bearing in this limit, $\partial/\partial y = 0$ and the gap narrows over $D$ from $h_0+d$ to $h_0$, and \Eqref{eq:pre:pressure} leads us to expect
\begin{equation}
  \frac{Ud}{D}
  \;\sim\; \Bigl(\,\frac{1}{D} \times \frac{h_0^3}{\eta} \frac{p}{D}\,\Bigr) \;+\; \Bigl(\,\frac{1}{L} \times \frac{h_0^3}{\eta} \frac{p}{L}\,\Bigr)\,. \label{eq:pre:sliderScale}
\end{equation}
With $D\ll L$, we drop the second term on the RHS: the Poiseuille flow is predominantly along $x$, \partFigref{fig:pre:types}{a}. So,
\begin{equation}
    p \sim \frac{\eta U Dd}{h_0^3}\,. \label{eq:pre:LongsliderPressure}
\end{equation}
This and shear stress $\sigma \sim \eta U/h_0$~\cite{Note1}
act on area $LD$ to give 
\begin{equation}
  \begin{split}
    &\load \sim \frac{\eta U d L D^2}{h_0^3}\,,\quad\drag \sim \frac{\eta U\! L D}{h_0}\\[3pt]
    &{}\qquad\Rightarrow\quad S \sim \frac{d^2}{D^2}\Bigl(\frac{h_0}{d}\Bigr)^3\,,\quad \mu \sim \frac{d}{D} \Bigl (\frac{h_0}{d} \Bigr)^2\\[6pt]
    &{}\hspace{10em}\Rightarrow\quad \mu \sim
    \Bigl(\frac{D}{d} \Bigr)^{1/3} S^{2/3}\,.
  \end{split}
  \label{eq:pre:longScalingDimension}
\end{equation}
The `anomalous exponent' of $2/3$ arises from $r_0= \sqrt{h_0R}$ in \Eqref{eq:pre:ballScale}, because a sphere is characterised by a single length scale $R$, while the $d$ and $D$ in \Eqref{eq:pre:LongsliderPressure} are constants unrelated to $h_0$, \ie\ a slider bearing is specified by {\it two} length scales. 

In the short-bearing limit ($L\ll D$), we drop the first term on the RHS of \Eqref{eq:pre:sliderScale}: volume conservation is now entirely managed by side-leakage~\cite{dubois1953analytical}, \partFigref{fig:pre:types}{b}, and we have
\begin{equation}
    p \sim \frac{\eta U d L^2}{Dh_0^3}\,. 
\end{equation}  
This and shear stress $\sigma \sim \eta U/h_0$ act on area $LD$ to give 
\begin{equation}
  \begin{split}
    &N \sim \eta U d L^3/h_0^3\,,\quad\drag \sim \frac{\eta U\! L D}{h_0}\\[3pt]
    &{}\qquad \Rightarrow\quad S \sim \frac{d^2}{L^2}\Bigl(\frac{h_0}{d}\Bigr)^3\,,\quad \mu \sim \frac{Dd}{L^2} \Bigl (\frac{h_0}{d} \Bigr)^2\\[6pt]
    &{}\hspace{10em}\Rightarrow\quad \mu \sim
    \Bigl(\frac{D^3}{L^2d} \Bigr)^{1/3} S^{2/3}.
  \end{split}
  \label{eq:pre:shortScalingDimension}
\end{equation}
The anomalous scaling therefore remains unchanged in the short-bearing limit, which only introduces new pre-factors. 

\subsection{Small-gap limit, $h_0 \lesssim d$ \label{sec:pre:scalingProfiles}}

Now, the gap profile matters. Figure~\ref{fig:pre:steps} (a)-(c) show increasing abruptness of the transition from $h(x=0) = h_0 + d$ to $h(D) = h_0$, with $\chi$ controlling the inlet length in (b) and (c).

\subsubsection{Wedge\label{sec:pre:scalingProfilesWedge}}

\begin{figure*}
    \centering
    \includegraphics{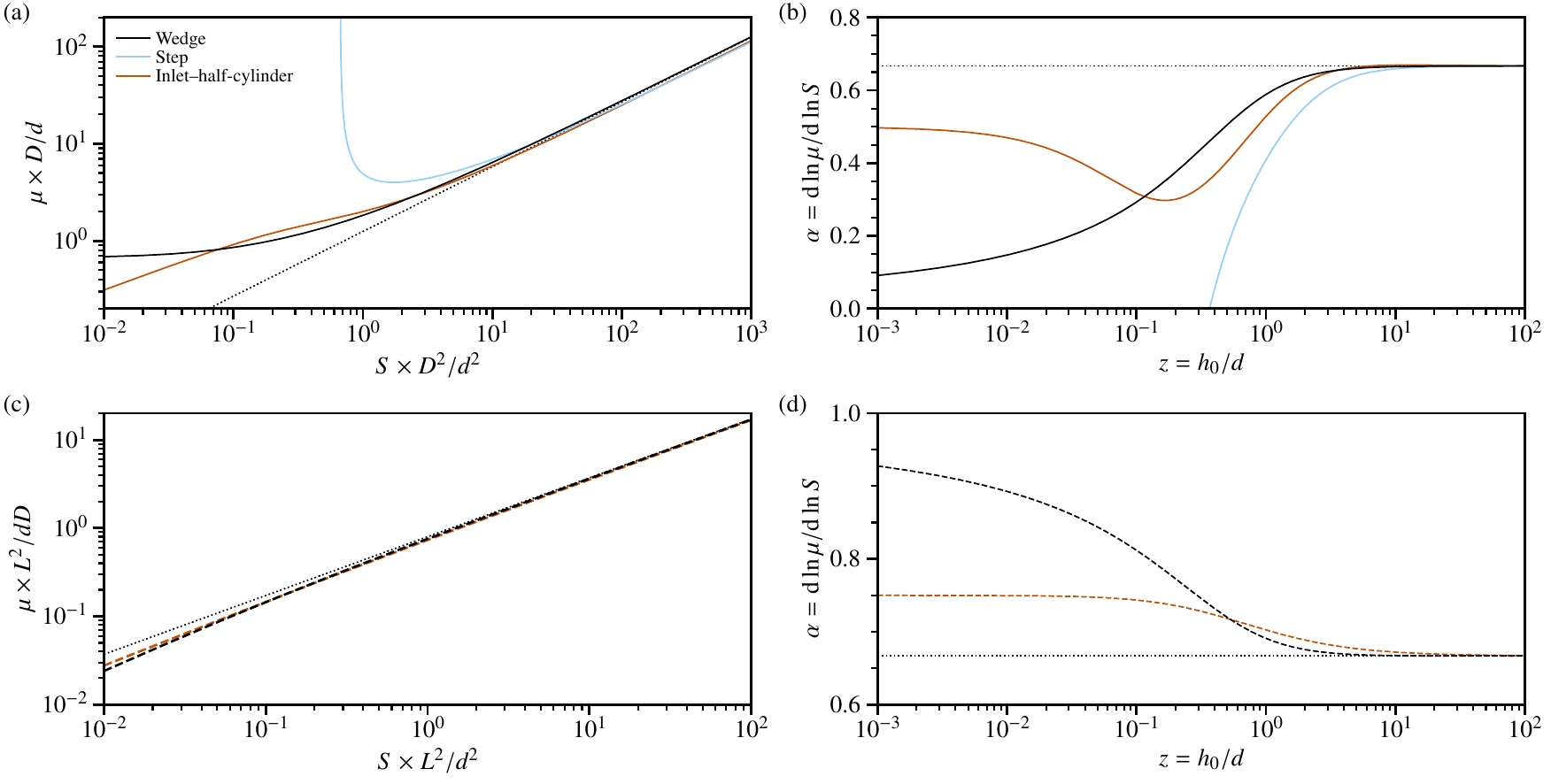}
    \caption{Stribeck curves, $\mu(S)$, for various gap profiles. (a)~Long-bearing limit, friction coefficient normalised by geometric pre-factors [$\mu/(d/D)$] as a function of Sommerfeld number scaled by geometric factor [$S/(d^2/D^2)$] from \Eqsref{eq:pre:Nlong} and \eqref{eq:pre:Flong}. High-$S$ scaling of $\mu \sim S^{2/3}$ shown by dotted line. Solid lines: black, wedge profile slider bearing, \Eqref{eq:pre:wedgeProfileDimension}; dark (orange), inlet--half-cylinder, \Eqref{eq:pre:inletProfile}; and, light (blue), Rayleigh step profile \Eqref{eq:pre:stepProfileDimension}. For step and inlet--half cylinder inlet length $\chi = 0.5$. (b) Running exponent $\alpha$ as a function of reduced gap, $h_0/d$. Lines as in (a). (c)~Short-bearing limit for same profiles [dashed lines, see legend in (a)] from \Eqsref{eq:pre:Nshort} and \eqref{eq:pre:Fshort}. Modified geometric pre-factors, $\mu/(dD/L^2)$ and $S/(d^2/L^2)$; $\chi=0.4$. (d) Respective running exponent.}
    \label{fig:pre:stribeck}
\end{figure*}

As $h_0 \to 0$ in a long wedge with profile 
\begin{equation}
  h=h_0+d \left( 1- \frac{x}{D}\right)\,,\label{eq:pre:wedgeProfileDimension}
\end{equation}
the contact becomes effectively non-conformal with a small-gap region of extent $x_0 \sim h_0 D/d$. From \Eqref{eq:pre:pressure} we expect
\begin{equation}
    \frac{1}{x_0} \times Uh_0 \sim \frac{1}{x_0} \frac{h_0^3}{\eta} \frac{p}{x_0}\quad \Rightarrow\quad p \sim \frac{\eta U x_0}{h_0^2}\,.
\end{equation}
This pressure and shear stress $\eta U/h_0$ acting over area $x_0L$ gives
\begin{equation}
    \load \sim \frac{\eta U L D^2}{d^2}\,,\quad \drag \sim \frac{\eta U D L}{d} \quad \Rightarrow\quad \mu \sim \frac{d}{D}\,,
\end{equation}
\ie\ a constant,~\partFigref{fig:pre:stribeck}{a} (black lines), with the running power law of the Stribeck curve (\ie\ the slope in a double-log plot), $\alpha = \rmd \ln \mu/\rmd \ln S \to 0$, \partFigref{fig:pre:stribeck}{b}. As $h_0 \to 0$, our problem becomes equivalent to `Taylor scraping flow'~\cite{batchelor2000introduction} with vanishing wedge angle, where there are logarithmic divergences in the forces; these  are missed in our scaling analysis, but will emerge in the full calculations in \Secref{sec:pre:specific}. 

For a short bearing, we again consider an effective non-conformal contact with $x_0 \sim h_0 D/d$. Equation~\eqref{eq:pre:pressure} now gives
\begin{equation}
    \frac{Uh_0}{x_0} \;\sim\; \Bigl(\,\frac{1}{x_0}\times\frac{h_0^3}{\eta} \frac{p}{x_0}\,\Bigr) \;+\; \Bigl(\,\frac{1}{L}\times\frac{h_0^3}{\eta} \frac{p}{L}\,\Bigr)\,. \label{eq:pre:ShortWedgeScaling}
\end{equation}
If we are only just into the small-gap limit ($h_0/d \lesssim 1$), then for a short bearing ($L \ll D$), $x_0/L = (h_0/d)(D/L) \gg 1$, and we can neglect the first term on the RHS of \Eqref{eq:pre:ShortWedgeScaling} and obtain
\begin{equation}
    p \sim \frac{\eta U L^2}{x_0 h_0^2}\,.
\end{equation}
This pressure and shear stress $\eta U/h_0$ acts on area $x_0L$ to give
\begin{equation}
  \begin{split}
  &N \sim p \times x_0L \sim \frac{\eta U L^3}{h_0^2}\,,\quad
  F \sim x_0 L \times \frac{\eta U}{h_0} \sim \frac{\eta U L D}{d}\\[3pt]
  &{}\qquad\Rightarrow\quad S = \frac{\eta UL}{N} \sim \left(\frac{h_0}{L}\right)^2, \quad \mu \sim \frac{Dh_0^2}{dL^2} \sim \left(\frac{D}{d}\right)S\,,
  \end{split}
\end{equation}
which is a linear scaling, \partFigref{fig:pre:stribeck}{c} and \partref{fig:pre:stribeck}{d}~(black dashed line).

Deep into the small gap regime, $h_0/d \lll 1$, and $x_0/L = (h_0/d)(D/L) \ll 1$ even for a short bearing ($D/L \gg 1$). Discarding the second term on the RHS of \Eqref{eq:pre:ShortWedgeScaling}, we find
\begin{equation}
    p \sim \frac{\eta U x_0}{h_0^2} \quad\Rightarrow\quad N \sim p \times x_0 L \sim \eta UL \left(\frac{D}{d}\right)^2. \label{eq:wedge:extraThin}
\end{equation}
With $\drag$ the same as before, we find
\begin{equation}
    \mu \sim \frac{d}{D}\,,
\end{equation}
as in the case of the long bearing in the same small-gap limit. 

\subsubsection{Inlet--half-cylinder\label{sec:pre:scalingProfilesInlet}}

Consistent with the assumptions underpinning \Eqref{eq:pre:pressure}, we describe the gap in the inlet--half-cylinder by a parabola, \partFigref{fig:pre:steps}{b},
\begin{equation}
  h=\Bigl\{
  \begin{array}{lll}
    h_0+d && 0\le x/D < \chi\,,\\
    h_0+d(1-x/D)^2/(1-\chi)^2 && \chi\le x/D\le 1\,,\label{eq:pre:inletProfileDimension}
  \end{array}
\end{equation}
with $0\le \chi\le1$. The low-$S$ limit for the long bearing is simple. For a small gap, the flat inlet region becomes inconsequential and we have a cylinder-on-flat geometry with $\mu \sim S^{1/2}$, \partFigref{fig:pre:stribeck}{a} and \partref{fig:pre:stribeck}{b}~[dark (orange)].

In the small-gap limit for a short bearing, leakage occurs sideways, so that we need only consider a small region near the bottom of the half cylinder ($x = D$ in our coordinates) where the gap remains of order $h_0$. Equation~\eqref{eq:pre:inletProfileDimension} shows that this region has dimension $x_0 = (1-\chi)D\sqrt{h_0/d}$. The rest of the scaling analysis proceeds along the same lines as that for the wedge in the thin-film limit, only with a different expression for $x_0$. 

Just into the thin-film regime, $x_0/L \gg 1$, we now find, \partFigref{fig:pre:stribeck}{b} and \partref{fig:pre:stribeck}{d}~[dashed dark (orange)],
\begin{equation}
    \mu \sim \frac{h_0}{L^2}(1-\chi)D\sqrt{\frac{h_0}{d}} \sim \left[ \frac{(1-\chi)D}{(Ld)^{1/2}}\right] S^{3/4}.
\end{equation}
Deep into the thin-film regime, $x_0/L \ll 1$, and again following the analysis for the wedge with a new expression for $x_0$ in \Eqref{eq:wedge:extraThin} we find $\mu \sim S^{1/2}$, with geometric pre-factors cancelling and the expected recovery of a long cylinder expression.

\subsubsection{Rayleigh step\label{sec:pre:scalingProfilesStep}}

The gap in the Rayleigh step is piecewise continuous,
\begin{equation}
  h=\Bigl\{
  \begin{array}{lll}
    h_0+d && 0\le x/D < \chi\,,\\
    h_0 && \chi\le x/D\le 1\,. \label{eq:pre:stepProfileDimension}
  \end{array}
\end{equation}
For a long-bearing with this profile in the small-gap limit ($h_0 \ll d$), the Couette flux in the thin gap region is negligible, and the compensating Poiseuile flux exists almost entirely to balance the Couette flux in the inlet of length $\chi D$ and height $d + h_0 \approx d$. From \Eqref{eq:pre:pressure}, we have
\begin{equation}
    \frac{Ud}{\chi D} \sim \frac{1}{\chi D} \times \frac{d^3}{12\eta}\times \frac{p}{\chi D} \quad\Rightarrow\quad p \sim \frac{\eta U \chi D}{d^2}\,.
\end{equation}
This acts over the whole length of the bearing (even though the flow in the narrow gap is negligible) to give the normal force, while the tangential load is dominated by the narrow gap:
\begin{equation}
    \load \sim  \frac{\eta U\! \chi D^2 L}{d^2}\,,\quad \drag \sim \frac{\eta U}{h_0} (1-\chi)DL\,,
\end{equation}
\ie, $N$ is a constant, while $F$ diverges as the gap narrows. So, we expect that as $S$ decreases, $\mu\sim F/N$ must reach a minimum and then upturns and diverges as $S=\eta UL/N$ approaches some constant value $\sim \chi^{-1} (d/D)^2$, \partFigref{fig:pre:stribeck}{a} and \partref{fig:pre:stribeck}{b}~[light (blue)].

In the short Rayleigh step, sideways Poisueille leakage occurs over the infinitesimal width of the step itself, so that the conditions necessary for the validity of the Reynolds lubrication formalism in \Eqref{eq:pre:pressure} no longer apply. 

\subsection{Summary of scaling analysis}

Our scaling analysis gives is a universal `anomalous' regime of $\mu \sim S^{2/3}$ at large $S$ for all three geometries in \Figref{fig:pre:steps}. When $h_0 \to d$, deviations from this scaling occurs. The details are geometry dependent, with the $S \to 0$ limit derivable by scaling analysis except in the pathological case of the short Rayleigh step. 
Solving \Eqref{eq:pre:pressure} by quadrature in Sections~\ref{sec:pre:exact} and \ref{sec:pre:alpha} will verify these scaling conclusions and provide closed forms for plotting the behaviours shown in \Figref{fig:pre:stribeck} and Table~\ref{tab:pre:limits} (and Fig.~3 of \Refcite{PRL}). Further, we will show that for a broad class of profiles, with small inlets, the deviation from $S^{2/3}$ scaling is upwards as $S$ decreases, which mimics entry into the BL regime even though lubrication remains hydrodynamic in origin. Understanding such deviation then allows, in \Secref{sec:data}, a general way of analysing experimental data to extract a surface texturing length scale from the Stribeck curve. 

\section{Exact lubrication calculations and limits\label{sec:pre:exact}}

\subsection{General lubrication calculations}

We start by deriving a general expression for the Reynolds lubrication pressure, $p(x, y)$  from \Eqref{eq:pre:pressure} in the long- and short-bearing limits, \Figref{fig:pre:types}a and b respectively. For completeness, we include some material that is available in textbooks~\cite{hamrock2004fundamentals}.

\subsubsection{Long slider bearings ($D\gg L$)\label{sec:pre:calcConfLong}}
With negligible edge effects and $p = p(x)$, $\partial_y \rightarrow 0$, and we can integrate \Eqref{eq:pre:pressure} with respect to $x$,
\begin{equation}
  \frac{\partial p}{\partial x}
  =6\eta U\Bigl(\frac{1}{h^2}-\frac{h_m}{h^3}\Bigr)\,,\label{eq:dpdx}
\end{equation}
where $h_m$ is a constant. Define a dimensionaless gap $\xi = h/d$, where $d$ is some fiducial length such as a `step' height, or the difference in height across the width of the bearing [$d = h(0) - h(D)$]. In terms of a dimensionless minimum gap $z = h_0/d$, the different bearing profiles are distinguished by $\delta(u) = \xi(u) - z$, where we have used the dimensionless length $u=x/D$ so that $0\le u\le 1$ spans the length of the bearing. Hereafter, the length-wise average $\langle \cdots \rangle$ is given by
\begin{equation}
  \uav{\cdots}=\frac{1}{D}\int_0^D\!\!\rmd x\,(\cdots)=\int_0^1\!\!\rmd u\,(\cdots)\,.\label{eq:uav}
\end{equation}
Integrating \Eqref{eq:dpdx} gives the pressure drop across the bearing
\begin{equation}
    \Delta p = 6\eta UD(\uav{\xi^{-2}}-\xi_m\uav{\xi^{-3}})/d^2\,.
\end{equation}
Assuming equal inlet and outlet pressures, $\Delta p=0$ and $\xi_m = \uav{\xi^{-2}}/\uav{\xi^{-3}}$. Integrating $p$ to obtain the normal and friction forces, \Eqref{eq:pre:loaddrag}, the transverse ($y$) integral trivially gives a factor of the width $L$.  We deal with the $x$ integral in the normal force by integrating \Eqref{eq:dpdx} by parts, $\int\! \rmd x\, p=-\int\!\rmd x\,x(\partial 
p/\partial x)$, again assuming $\Delta p = 0$. After some algebra, we find
\begin{gather}
    \load =\frac{6\eta U\!LD^2}{d^2}\,
    \frac{\uav{u\xi^{-3}}\uav{\xi^{-2}}-\uav{u\xi^{-2}}\uav{\xi^{-3}}}%
    {\uav{\xi^{-3}}}
    =\frac{\eta U\!LD^2}{d^2}g_{\delta}(z)\,,\label{eq:pre:Nlong}
    \\
        \drag = \frac{\eta U\!LD}{d}\,\frac{4\uav{\xi^{-1}}\uav{\xi^{-3}}-3\uav{\xi^{-2}}^2}%
        {\uav{\xi^{-3}}}
        = \frac{\eta U\! L D}{d}f_{\delta}(z)\,,\label{eq:pre:Flong}
\end{gather}
where the subscript in the dimensionless normal load, $g(z)$, and drag, $f(z)$, points to the dependence on the gap profile $\delta(u) = \xi(u)-z$. These expressions imply $\mu = F/N \sim d/D$ and $S = \eta UL/N \sim d^2/D^2$, reproducing the geometric factors in \Eqref{eq:pre:longScalingDimension} established by scaling analysis. 

The Stribeck curve, $\mu(S)$, then follows parametrically as the gap $z$ (and so $\uav{\xi^{-n}}$) varies as a control parameter. Note that the higher-order averages required in \Eqsref{eq:pre:Nlong} and \eqref{eq:pre:Flong} can be computed from $\uav{\xi^{-1}}$ and $\uav{u\,\xi^{-1}}$ from recurrence:
\begin{equation}
  \uav{\xi^{-n-1}}=-\frac{1}{n}
  \frac{\leibnizd{\uav{\xi^{-n}}}}{\leibnizd{z}}\,,\quad
    \uav{u\,\xi^{-n-1}}=-\frac{1}{n}
    \frac{\leibnizd{\uav{u\,\xi^{-n}}}}{\leibnizd{z}}\,,\label{eq:recur}
\end{equation}
which follow by differentiating \Eqref{eq:uav} and using $\xi=z+\delta$.

\subsubsection{Short slider bearings ($D \ll L$)\label{sec:pre:calcConfShort}}

In this limit, we neglect $\partial p/\partial x$ in \Eqref{eq:pre:pressure} and, recalling that $h$ is $y$-independent, the equation simplifies to
\begin{equation}
 h^3\frac{\partial^2 p}{\partial y^2}
  =6\eta U\frac{\rmd h}{\rmd x}\,.
\end{equation}
Fixing $p(y=0)=p(y=L)=0$, \ie\ taking $p$ as the excess pressure, this integrates to
\begin{equation}
p = - \frac{3\eta U}{h^3}\frac{\rmd h}{\rmd x}\times y(L-y)\,,
\end{equation}
giving the normal load
\begin{equation}
    \begin{split}
  \load = 3\eta U\,\frac{L^3}{6} \int_0^1\!\frac{\rmd u}{\xi^3}\,\frac{\rmd \xi}{\rmd u} 
   =\frac{\eta U\!L^3}{d^2}\,\frac{(2z+1)}{4z^2(z+1)^2}
   = \frac{\eta U \! L^3}{d^2}g(z)\,.\label{eq:pre:Nshort}
  \end{split}
\end{equation}
Now, $g(z)$ does not depend on the bearing profile, $\delta$, but only the dimensionless minimum gap, $z$.  The drag also simplifies due to the absence of Poiseuille flow along the sliding direction,
\begin{equation}
  \drag=L\int_0^D\!\!dx\,\frac{\eta U}{h}=\frac{\eta U\!L D}{d}\,    f_{\delta}(z)\,,\label{eq:pre:Fshort}
\end{equation}
with $f_{\delta}(z) = \uav{\xi^{-1}}$.
From \Eqsref{eq:pre:Nshort} and \eqref{eq:pre:Fshort}, $S$ and $\mu$ can again be derived with the geometric pre-factors previously established in the scaling argument, \Eqref{eq:pre:shortScalingDimension}. 

\subsubsection{Power-law expansion\label{sec:pre:powerExpand}}

In the large-gap limit, $z \gg 1$. For the short bearing, expanding \Eqref{eq:pre:Nshort} gives $g(z) \sim z^{-3}$. For the long bearing, we need to expand the $\xi$-dependent numerators in \Eqsref{eq:pre:Nlong} in the small quantity $\delta(u)/z \sim \order{1/z}$, whereupon we find
\begin{align}
    \uav{\xi^{-n}} &= \int_0^1\! \frac{\rmd u\, z^{-n}}{[1 + \delta(u)/z]^{-n}} \approx z^{-n}(1-n\uav{\delta}/z)\,,\\[3pt]
    \uav{u\xi^{-n}} &= \int_0^1\! \frac{\rmd u\, z^{-n}u}{[1 + \delta(u)/z]^{-n}} \approx z^{-n}(1/2-n\uav{u\delta}/z)\,.
\end{align}
The leading order terms cancel, leaving
\begin{equation}
    g(z) \approx \frac{6z^{-5}(\uav{\delta}/2z-\uav{u\delta}/z}{z^{-3}} = \frac{3(\uav{\delta}-2\uav{u\delta})}{z^{3}} \sim \order{\frac{1}{z^3}}\,,
    \label{eq:pre:delta}
\end{equation}
as in the short-bearing limit. In the final step we have taken the gap profile averages to be of order the `step' height and not cancel, which is correct for all profiles apart from the case of $\chi = 1$, which violates the lubrication approximation. With the tangential load $f(z) \sim z$ in all cases (as $\uav{\xi^{-n}}\rightarrow z^{-n}$), we recover $S \propto z^3$ with $\mu \propto z^2 \propto S^{2/3}$, as previously found.

In the small-gap limit, $z+1 \rightarrow 1$ and $g(z) = 1/4z^2 \propto z^{-2}$ from \Eqsref{eq:pre:Nlong} and \eqref{eq:pre:Nshort}. This gives $S \propto z^2$ and $\mu \propto z$, \ie\ $\mu \propto S^{1/2}$ for all bearing profiles and types, contradicting our scaling predictions. A more rigorous analysis is needed.

\subsection{Specific gap profiles\label{sec:pre:specific}}

\begin{table}[t]
\begin{ruledtabular}
  \begin{tabular}{rccccc}
    Profile & $g(z) \propto N$ & $f(z)\propto F$ & $S(z)$ & $\mu(z)$ & $\mu(S)$ \\[3pt]
    \hline\\[-3pt]
     \vtop{\hbox{\strut Wedge: Long}}
    &  $ 6\ln z^{-1}$ &  $4\ln z^{-1}$
    &  $\textstyle\frac{1}{6\ln z^{-1}}$
    &  $\textstyle\frac{2}{3}$ & $\sim S^{0}$\\[4pt]
     \vtop{\hbox{\strut Short}}
    &  $\textstyle \frac{1}{4z^2}$ & $\ln z^{-1}$
    &  $4z^2$
    &  $4z^2\ln z^{-1}$ & $\sim S^{1}$\\[4pt]
     \vtop{\hbox{\strut IHC: Long}}
    &  $\textstyle \frac{3(1-\chi)^2}{z}$ &  $\textstyle \frac{\pi(1-\chi)}{z^{1/2}}$
    &  $\textstyle \frac{z}{3(1-\chi)^2}$
    &  $\textstyle \frac{\pi z^{1/2}}{3(1-\chi)}$ & $\sim S^{1/2}$\\[4pt]
     \vtop{\hbox{\strut Short}}
    &  $\textstyle \frac{1}{4z^2}$ &  $\textstyle\frac{\pi(1-\chi)}{2\sqrt{z}}$
    &  $4z^2$
    &  $2\pi(1-\chi)z^{3/2}$ & $\sim S^{3/4}$\\[4pt]
     \vtop{\hbox{\strut Step: Long}}
    &  $3 \chi$ &  $\textstyle \frac{1-\chi}{z}$
    &  $\textstyle\frac{1}{3\chi}$
    &  $\textstyle\frac{1-\chi}{3\chi z}$ & $\to \infty $\\[2pt]
  \end{tabular}
\caption{Small-gap scaling behaviour, $z \ll 1$, for varying gap profiles, $\delta(u)$, in both the long and short bearing limits (first column) for wedge, inlet--half-cylinder (IHC) and step. Scaling is given for the normalised normal load ($g$), tangential load ($f$), Sommerfeld number ($S$) and friction coefficient ($\mu$) in terms of the minimum gap, respectively. The final column is the resultant low-$z$ scaling of the Stribeck curve, $\mu(S)$.\label{tab:pre:limits}}
\end{ruledtabular}
\end{table}

For each gap profile, full expressions for the load and drag in the long, \Eqsref{eq:pre:Nlong}--\eqref{eq:pre:Flong}, and short, \Eqsref{eq:pre:Nshort}--\eqref{eq:pre:Fshort}, bearing limits can be calculated, see Appendix, giving the Stribeck curves in \Figref{fig:pre:longSteps} and \ref{fig:pre:shortSteps} respectively, and small-gap scaling limits in Table~\ref{tab:pre:limits}. For all step types and most inlet length length, we indeed find (see Appendix) $\mathcal{O}(1)$ pre-factors. The physical picture underlying our scaling analysis therefore holds. Such analysis can go further: it can treat a changing Poiseuille flow direction with reducing gap and predicts the recovery of the long bearing limit for short bearings for $h_0 \lll d$. 

However, one discrepancy highlights an important limitation of our narrow-gap scaling approach, which gives a constant $\mu$ as $h_0 \rightarrow 0$ because both $\load$ and $\drag$ individually approach constant values. From quadrature, we see that $N$ and $F$ both diverge logarithmically, Table~\ref{tab:pre:limits}, but these cancel to give a constant $\mu$. In considering only a narrow-gap region with $h\sim h_0$, the cumulative effects of the region with a larger gap are neglected, as the Poiseuille pressure drop and Couette drag fall off with increasing $h_0$. Where the gap increases quadratically this assumption is correct~\cite{warren2016sliding}. However, it is not strictly correct for a linear profile. 

\begin{figure}[t]
    \centering
    \includegraphics{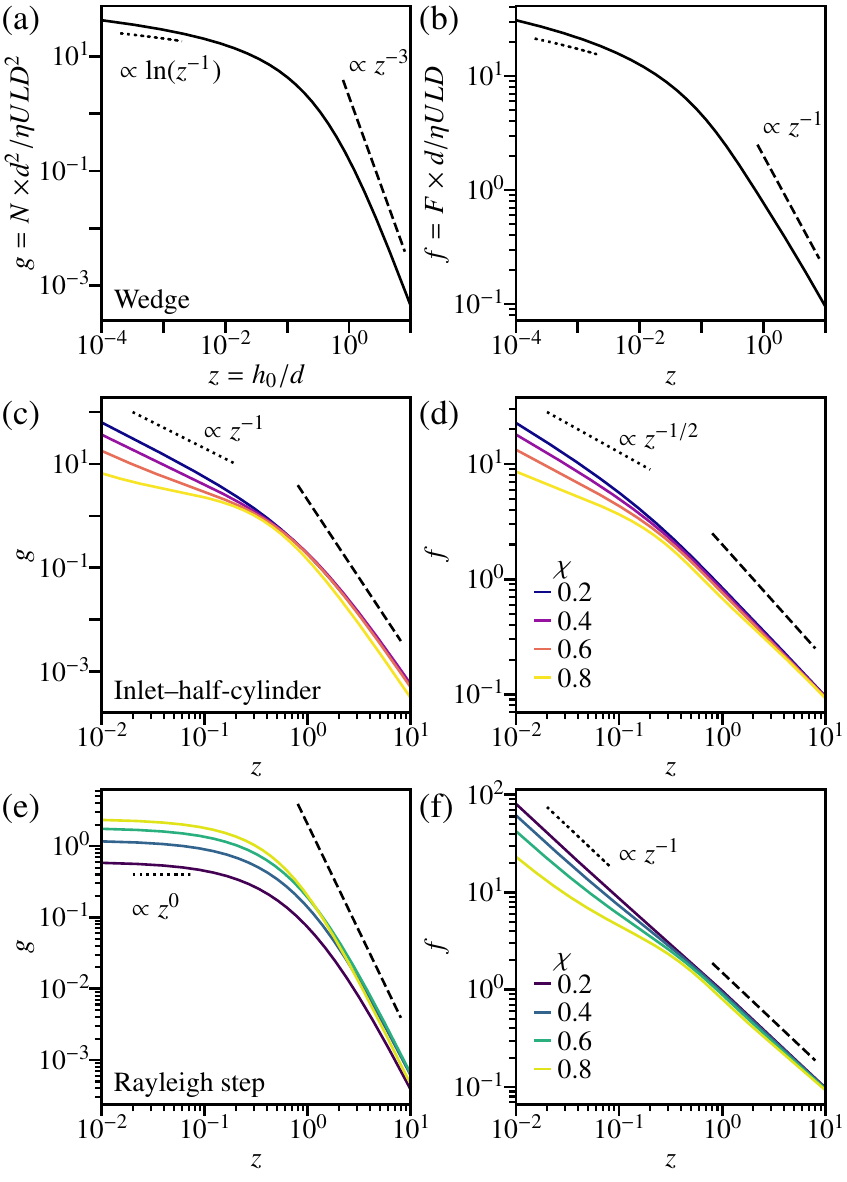}
    \caption{Load and drag dependence with gap height for long-bearing conformal `steps'. Dimensionless load, $g = \load/(\eta U L D^2/d^2)$, and drag, $f = \drag/(\eta U L D/d)$, as a function of reduced gap, $z = h_0/d$. (a)~Wedge profile load (solid line) with large gap limit applying to all profiles, $\load \propto z^{-3}$ (dashed line), and profile-specific low-$z$ limit, $\load \propto \ln(z^{-1})$ (dotted line). (b)~corresponding drag with high-$z$ limit, $\drag \propto z^{-2}$ (dashed line), and low-$z$ limit, $\drag \propto \ln(z^{-1})$ (dotted line). (c)~Load for inlet--half-cylinder with increasing inlet length, $\chi$ (dark to light), see legend in (d) for values. (d)~Corresponding drag. (e)~Load for Rayleigh step profile with increasing $\chi$, see (f) for values, and low-$z$ limit $\propto z^{-1}$. (f)~Corresponding drag.}
    \label{fig:pre:longSteps}
\end{figure}

\begin{figure}
    \centering
    \includegraphics{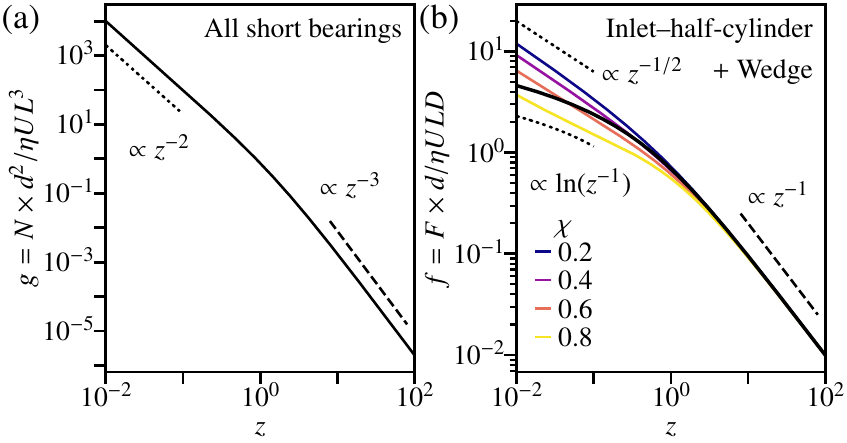}
    \caption{Short bearing load and drag gap dependence with changing step profile. (a)~Dimensionless load, $g(z) = \load/(\eta U L^3/d^2)$, as a function of reduced gap, $z=h_0/d$, for all short-bearing profiles. Scaling in large-gap limit $z \gg 1$, $g \propto z^{-3}$ (dashed line), and scaling in limit of $z \ll 1$, $g \propto z^{-2}$ (dotted line). (b)~Dimensionless drag, $f(z) = F/(\eta U L D/d)$, for wedge gap profile, \Eqref{eq:pre:wedgeProfile}, bold (black) line. Scaling for $z \gg 0$, $f \propto z^{-1}$ applies to all profiles (dashed lines); scaling for $z\ll 1$, $f \propto \ln z^{-1}$ (dotted line). Inlet--half-cylinder profile, \Eqref{eq:pre:inletProfile}, with increasing $\chi$, dark (blue) to light (gold) fine lines. Scaling for $z \ll 1$, $f \propto z^{-1/2}$. See Table~\ref{tab:pre:limits} for resulting scaling of Stribeck curve $\mu(S)$ for $z \to 0$.}
    \label{fig:pre:shortSteps}
\end{figure}

\section{Transition from large-gap scaling\label{sec:pre:alpha}}

To probe how deviations from $\mu \sim S^{2/3}$ scaling first emerges as $h_0 \to d$, which will be important for analysing experiments, consider the `running exponent' of the Stribeck curve, 
\begin{eqnarray}
  \alpha &=& \frac{\leibnizd{\,\ln\mu}}{\leibnizd{\,\ln S}} =1-\frac{\leibnizd{\,\ln F}}{\leibnizd{\,\ln N}} = 1-\frac{\leibnizd{\,\ln f}}{\leibnizd{\,\ln g}}
   \nonumber \\
    &=& 1-\frac{\leibnizd{\,\ln f}}{\rmd z} \left( \frac{\leibnizd{\,\ln g}}{\rmd z}\right)^{-1}
    \!\!= 1 - \frac{1}{f}\frac{\rmd f}{\rmd z} \left(\frac{1}{g}\frac{\rmd g}{\rmd z}\right)^{-1}\!\!.
  \label{eq:pre:alpha3}
\end{eqnarray}

\subsection{Short-bearing limit\label{sec:pre:alpha_short}}

\begin{figure}
    \centering
    \includegraphics{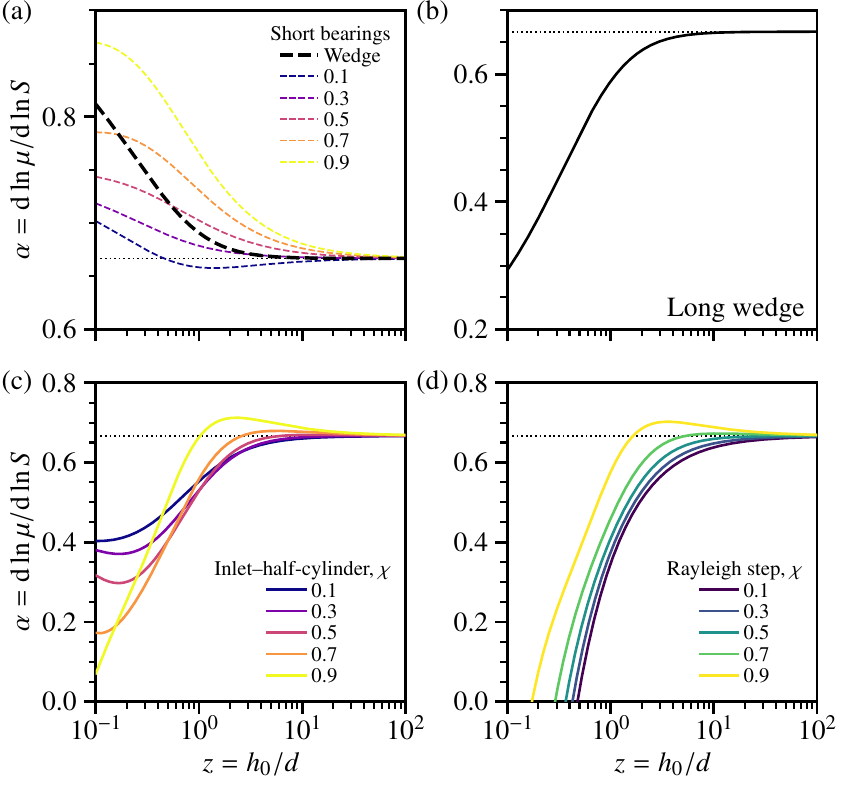}
    \caption{Running exponent $\alpha = \rmd \ln \mu / \rmd \ln S$ vs~gap around $z \approx 1$. $\alpha < 2/3$ corresponding to upwards deviations in Stribeck curve. (a)~Deviation from large-gap scaling (dotted line) for short-bearing limit. Dashed lines: bold (black), wedge; fine, inlet--half-cylinder with increasing inlet, $\chi$, dark (purple) to light (yellow), see legend for values. (b)~Long-bearing limit wedge. (c)~Long inlet--half-cylinder with increasing inlet length [dark (purple) to light (yellow)], see legend for values. (d)~Long Rayleigh step with increasing inlet length dark (purple) to light (gold), see legend for values.  Rayleigh step; and, dark (orange), inlet--half-cylinder. Calculated from \Eqref{eq:pre:alpha3} with $f$ and $g$ given in Appendix.}
    \label{fig:pre:alpha}
\end{figure}

Given $g(z)$ for all profiles, \Eqref{eq:pre:Nshort}, and the recurrence relation for the derivative of $f(z)$, \Eqref{eq:recur}, we find
\begin{equation}
  {\alpha}=1-\frac{z(z+1)(2z+1)}{2(3z^2+3z+1)}\times
  \frac{\uav{\xi^{-2}}}{\uav{\xi^{-1}}}\,.
\end{equation}
The moments of $\xi$ can be recast explicitly in terms of the gap profile, $\xi(u) = z + \delta(u)$. In the asymptotic limit of large $z$,
\begin{equation}
  \biguav{\frac{1}{\xi^n}}=\frac{1}{z^n}
  \Bigl[1-\frac{n\uav{\delta}}{z}
    +\mathcal{O}\Bigl(\frac{1}{z^2}\Bigr)\Bigr]\,.
\end{equation}
Then, within the short bearing approximation,
\begin{equation}
  \alpha=\frac{2}{3}+\frac{\uav{\delta}-1/2}{3z}
  + \mathcal{O}\Bigl(\frac{1}{z^2}\Bigr)\,.
\end{equation}
The sign of initial deviations from $\alpha = 2/3$ depends on the average profile height $\uav{\delta}$. The short inlet--half-cylinder is `blunt', $\uav{\delta} < 1/2$, only when $\chi$ is small enough, \partFigref{fig:pre:alpha}{a} [light (gold) to dark (blue) for decreasing $\chi$], in which case $\alpha$ initially drops below 2/3 and the Stribeck curve bends upwards from the region of universal $S^{2/3}$ scaling, mimicking the onset of EHL. The onset of deviations from $\alpha = 2/3$ for the short wedge, \partFigref{fig:pre:alpha}{a} (black dashed), is at significantly smaller $z$ than for the inlet--half-cylinder because $\uav{\delta} = 1/2$, so that deviations, which are always positive, scale as $z^{-2}$.  

\subsection{Long-bearing limit\label{sec:pre:alpha_long}}
In the long-bearing limit, we find a more complex expression,
\begin{equation}
  \alpha=\frac{2}{3}+\frac{2(\uav{u}\uav{\delta^2}-\uav{u\delta}\uav{\delta})
    +\uav{u\delta}^2-\uav{u}\uav{\delta^2}}
  {3d(\uav{u}\uav{\delta}-\uav{u\delta})z}
  {}+\mathcal{O}\Bigl(\frac{1}{z^2}\Bigr)\,.\label{eq:pre:longTermGeneral}
\end{equation}
To highlight the structure of the answer we have retained $\uav{u}=1/2$, rather than substitute its value. The sign of the initial deviation from $\alpha = 2/3$ is therefore controlled by the numerator of the second term in this expansion~\cite{Note2}. 
For the wedge, this reduces to
\begin{equation}
  \alpha=\frac{2}{3}-\frac{17}{90z^2}
  +\mathcal{O}\Bigr(\frac{1}{z^3}\Bigl)\,.
\end{equation}
So, $\alpha$ drops below 2/3, \partFigref{fig:pre:alpha}{b}, leading to an upwards deviation of the Stribeck curve from $S^{2/3}$ scaling. Again, this effect is second order in $1/z$, as in the short-bearing case.

For the inlet--half-cylinder,
\begin{equation}
  \alpha=\frac{2}{3}-\frac{2(1+5\chi-30\chi^2)}{45(1+3\chi)}\frac{1}{z}
  +\mathcal{O}\Bigl(\frac{1}{z^2}\Bigr)\,.
\end{equation}
For sufficiently short inlet lengths, \ie\ $\chi < (5 + \sqrt{145}) / 60 \approx 0.284$, in this narrow-gap limit, $\alpha$ is lowered, \partFigref{fig:pre:alpha}{c}~(dark lines), and the Stribeck curve deviates upwards from the large-gap scaling with decreasing $S$. 

Finally, for the Rayleigh step, \partFigref{fig:pre:alpha}{d}, we find
\begin{equation}
  \alpha=\frac{2}{3}-\frac{1-2\chi}{3z}
  +\mathcal{O}\Bigr(\frac{1}{z^2}\Bigl)\,
  ,
\end{equation}
so that in a similar manner, a suitably `blunt' step, \ie\ $\uav{\delta} = \chi < 1/2$, will give $\alpha < 2/3$.

We therefore arrive at a conclusion for all slider bearing types and profiles: large-gap scaling is $\mu \sim S^2/3$ and upwards deviation with decreasing $S$ as the gap approaches the step height requires a small enough $\uav{\delta}$, \ie\ a small enough inlet.

\section{Interpretation of experimental data\label{sec:data}}

We have shown that for a broad class of surface profiles and bearing dimensions, upwards deviation from $\mu \sim S^{2/3}$ scaling in the HL regime starts when the gap approaches the step height, \ie\ $h_0/d \sim 1$. This also apply to various textured surfaces, \Figref{fig:pre:steps}(a), so that $d$ can be estimated from measured Stribeck curves. We end by giving a worked example of this procedure.

\begin{figure}
    \centering
    \includegraphics{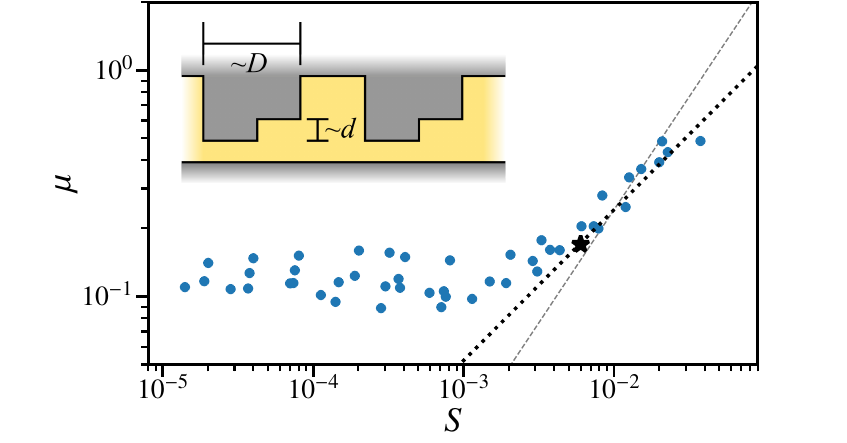}
    \caption{Stribeck curve for molten chocolate lubricating a bio-mimetic tongue sliding on steel. Re-plotted from \Refcite{soltanahmadi_insights_2023} as $\mu$ against $S = \eta U D/N$ using $N = \SI{1.25}{\milli\newton}$ (see text). Lines: bold dotted, $S^{2/3}$ scaling; fine dashed, $\mu \sim S$. Inset schematic: relevant length scales, $D \sim \SI{350}{\micro\meter}$ from papillae diameter, and $d \sim \SIrange{30}{60}{\micro\meter}$ from papillae roughness.}
    \label{fig:tongue}
\end{figure}

Soltanahmadi \etal~\cite{soltanahmadi_insights_2023} obtained the Stribeck curves for a range of geometries lubricated by molten chocolate. We show their data for a bio-mimetic tongue against a steel surface, \Figref{fig:tongue}. The $\SI{2}{\centi\meter} \times \SI{2}{\centi\meter}$ `tongue' surface consists of $80\times \SI{500}{\micro\metre}$ high hemispherical `fungiform' papillae and $800 \times \SI{250}{\micro\metre}$ high cylindrical `filiform' papillae, each of which is textured on the \SIrange{10}{100}{\micro\metre} scale (see Fig.~3a in~\Refcite{andablo-reyes_3d_2020}). Their material parameters and applied normal load of $\SI{1}{\newton}$ implies that the fungiform papillae will compress $\approx \SI{200}{\micro\metre}$ and bring the more numerous filiform papillae into proximity with the lower surface at a gap height comparable to the papillae texturing. 

In the large gap HL regime, where all papillae have comparable gap heights and the more numerous papilla form dominates, the data presented in Fig.~5B(2) of~\Refcite{soltanahmadi_insights_2023} can be replotted as the Stribeck curve for a single filiform papilla at a normal load of $1/800\, \si{\newton}$, \Figref{fig:tongue}. In the high-$S$ regime, the scaling law slope is less than unity (fine dashed line), but is consistent with $\mu \sim S^{2/3}$ (dashed line); fitting the data at $S > 10^{-3}$ gives $\alpha = 0.6(1)$. This regime can therefore be interpreted in terms of our analysis of the HL of a step slider bearing. 

The properties of the effective step bearing can be estimated from the point ($\mu^\star \simeq 0.17,~S^{\star} \simeq 6\times10^{-3}$) at which the Stribeck curve deviates from $S^{2/3}$ scaling, \Figref{fig:tongue} (black star). For a long bearing,  \Eqref{eq:pre:longScalingDimension} shows that when $h_0 = d$, we have
\begin{equation}
    S^{\star} \sim \frac{d^2}{D^2}\,,\quad \mu^{\star} \sim \frac{d}{D}\, \quad\Rightarrow\quad \sqrt{S^{\star}} \sim \mu^\star\,.
\end{equation}
With an observed $S^\star \simeq 0.006$, our scaling analysis of the HL regime of a long slider bearing therefore predicts $\mu^\star \sim \sqrt{S^\star} = 0.08$, which is order-of-magnitude consistent with the observed $\mu^{\star} \simeq 0.17$~\cite{Note3}
Taking $D = \SI{350}{\micro\metre}$ (the diameter of the cylindrical papillae) then gives an estimate for $d$, either from $D\sqrt{S^{\star}} \simeq \SI{30}{\micro\metre}$ or from $D\mu^{\star} \simeq \SI{60}{\micro\metre}$, consistent with the actual papillae roughness on the \SIrange{30}{60}{\micro\meter} scale~\cite{andablo-reyes_3d_2020}.

Interestingly, Soltanahmadi \etal~\cite{soltanahmadi_insights_2023} estimated the gap height in the vicinity of $(S^\star,\mu^\star)$ to be $h_0 \gtrsim \SI{e0}{\micro\meter}$ [see their Fig.~5B(1)] using a phenomenological expression~\cite{Myant2010} modified from an analytical form derived from an analysis of EHL~\cite{deVicente2005}, which is significantly smaller than the gap size we have estimated based on our HL analysis. However, the actual measured gap size is  $\simeq \SI{50}{\micro\metre}$ (see \Refcite{soltanahmadi_insights_2023} Supporting Information Fig.~S8B), which is well within range of our estimate based on the point of deviation from $S^{2/3}$ scaling being where $h_0 \simeq d$, where  $d \simeq \SIrange{50}{100}{\micro\meter}$ is the scale of surface texturing.  

\section{Conclusions\label{sec:pre:conc}}

We have shown that the HL of non-conformal and conformal contacts differ fundamentally. For former, the problem has a single length scale, the minimum gap, giving `trivial'  scaling in the Stribeck curve, \eg\ $\mu \sim S$ for ball-on-flat, independent of the absolute gap. In contrast, the HL of conformal contacts is non-trivial due to the presence of two independent length scales, the minimum gap and a surface texturing or `step' height, giving anomalous $\mu\sim S^{2/3}$ large-gap scaling. As the gap decreases below the step height, deviation from $S^{2/3}$ occurs. `Blunt' surface profiles, where the gap remains small over more of the bearing area, show an `upwards' deviation as $S$ drops, reminiscent of entry into the EHL regime. We used our approach to deduce the length scale of surface texturing from the data in a recent triborheological study.

Large-gap HL is of limited application in engineering applications, which are typically designed to function near the minimum of the Stribeck curve. However, this regime becomes important whenever soft matter is involved in lubrication, \eg, in the oral processing of food~\cite{soltanahmadi_insights_2023}, the application of skin cream~\cite{lee2022predictive}, and the extrusion of ceramic `green bodies'~\cite{jiang2009effect} (a ceramic `green body' is the formed object before firing). Future work may therefore fruitfully extend our analysis to non-Newtonian fluids with, \eg, rate-dependent rheology and normal stress differences.

\appendix
\setcounter{equation}{0}
\renewcommand\theequation{A\arabic{equation}}
\section*{Appendix: gap profile calculations}\label{app:profiles}

\subsection{Wedge}

In dimensionless form the gap, \Eqref{eq:pre:wedgeProfileDimension}, of the wedge is
\begin{equation}
  \xi=z+1-u\,.\label{eq:pre:wedgeProfile}
\end{equation}
The averages required for \Eqsref{eq:pre:Nlong} and \eqref{eq:pre:Flong} then evaluate to
\begin{equation}
\begin{split}
  &\textstyle\uav{\xi^{-1}} = \ln\frac{z+1}{z}\,,\quad
  \uav{\xi^{-2}} = \frac{1}{z(1+z)}\,,\quad
  \uav{\xi^{-3}} = \frac{1+2z}{2z^2(1+z)^2}\,,\\[3pt]
  &\textstyle\uav{u\,\xi^{-2}} = \frac{1}{z} - \ln\frac{z+1}{z}\,,\quad
  \uav{u\,\xi^{-3}} = \frac{1}{2z^2(1+z)}\,. 
\end{split}
\end{equation}
In the long-bearing limit, the dimensionless functions are
\begin{equation}
  f=4\ln\frac{1+z}{z}-\frac{6}{1+2z}\,,\quad
  g=6\ln\frac{1+z}{z}-\frac{12}{1+2z}\,.
\end{equation}
In the limit of $z \to 0$, \ie\ with $z + 1 \to 1$, 
\begin{equation}
      \lim_{z \to 0} \{f, g\} = \{6\ln z^{-1}, 4 \ln z^{-1}\}
      \;\Rightarrow\;
    \lim_{z \to 0} \mu = \frac{2}{3}\,\frac{d}{D}\,.
\end{equation}
In the short bearing limit, $f = \uav{\xi^{-1}} \rightarrow \ln z^{-1}$ follows trivially alongside $g(z) \rightarrow 1/4z^2$, as derived in \Secref{sec:pre:powerExpand}.

\subsection{Inlet--half-cylinder}

Recasting the profile, \Eqref{eq:pre:inletProfileDimension}, in dimensionless terms,
\begin{equation}
  \xi=\Bigl\{
  \begin{array}{lll}
    z+1 && 0\le u \le \chi\,,\\
    z+(1-u)^2/(1-\chi)^2 && \chi\le u\le 1\,.\label{eq:pre:inletProfile}
  \end{array}
\end{equation}
The averages then evaluate to
\begin{equation}
\begin{split}
  \uav{\xi^{-1}} & = \textstyle\frac{\chi}{1+z}+\fubar{1-\chi}\,,\\
  \uav{\xi^{-2}} & = \textstyle\frac{\chi}{(1+z)^2}    + \frac{1-\chi}{2z}  \Bigl(\frac{1}{1+z}               + \fubar{1}\Bigr)\,,\\
  \uav{\xi^{-3}} & = \textstyle\frac{\chi}{(1+z)^3}    + \frac{1-\chi}{8z^2}\Bigl(\frac{3+5z}{(1+z)^2}         + \fubar{3}\Bigr)\,,\\
  \uav{u\,\xi^{-2}} & = \textstyle\frac{\chi^2}{2(1+z)^2} + \frac{1-\chi}{2z}  \Bigl(\frac{\chi}{1+z}                + \fubar{1}\Bigr)\,,\\
  \uav{u\,\xi^{-3}} & = \textstyle\frac{\chi^2}{2(1+z)^3} + \frac{1-\chi}{8z^2}\Bigl(\frac{1+z+2\chi(1+2z)}{(1+z)^2} + \fubar{3}\Bigr)\,.
\end{split} \label{eq:pre:inletxi}
\end{equation}
These expressions used in \Eqsref{eq:pre:Nlong} and \eqref{eq:pre:Flong} (long bearing) or \Eqsref{eq:pre:Nshort} and \eqref{eq:pre:Fshort} (short bearing) parametrically gives \Figref{fig:pre:stribeck}~[dark (orange) lines]. We do not show the lengthy expressions for $f(z)$ and $g(z)$ in this case.

For long bearings, a more rigorous analysis of the limits of \Eqref{eq:pre:inletxi} with $\tan^{-1}(z^{-1/2})/\sqrt{z} \to \pi/2\sqrt{z} - 1$ gives $g(z) \to 3(1-\chi)^2/z$, \partFigref{fig:pre:longSteps}{c}, and $f \to \pi(1-\chi)/\sqrt{z}$, \partFigref{fig:pre:longSteps}{d}. The $1- \chi$ originates from a shorter inlet making a larger cylinder, $R = D^2(1-\chi)^2/2d$. The resultant $S$ and $\mu$ are given in Table~\ref{tab:pre:limits}.

For short bearings, taking the limits of \Eqref{eq:pre:inletxi} gives $f(z) \to 2\pi(1-\chi)z^{3/2}$, \partFigref{fig:pre:shortSteps}{b}. With the $g(z)$ for all short bearings, we recover $\mu \propto S^{3/4}$ scaling with included $\order{1}$ numerical prefactor for $\chi \lesssim 0.9$, see Table~\ref{tab:pre:limits}.

\subsection{Step}

In non-dimensional form, \Eqref{eq:pre:stepProfileDimension} becomes
\begin{equation}
  \xi=\Bigl\{
  \begin{array}{lll}
    z+1 && 0\le u \le \chi\,,\\
    z && \chi\le u\le 1\,. \label{eq:pre:stepProfile}
  \end{array}
\end{equation}
The gap averages to calculate the $\load$ and $\drag$, and hence $\mu(S)$, for this problem are then, rather trivially,
\begin{equation}
  \textstyle
  \uav{\xi^{-n}} = \frac{\chi}{(1+z)^n}+\frac{1-\chi}{z^n}\,,\quad
  \uav{u\,\xi^{-n}} = \frac{\chi^2}{2(1+z)^n}+\frac{1-\chi^2}{2z^n}\,.\label{eq:pre:stepxi}
\end{equation}
For the Rayleigh step in the long-bearing limit the dimensionless functions, \partFigref{fig:pre:longSteps}{f}~(dark to light with increasing $\chi$), are
\begin{equation}
  \begin{split}
    \textstyle
    f=\frac{3\chi(1-\chi)}{\chi z^3+(1-\chi)(1+z)^3}+\frac{\chi}{1+z}+\frac{1-\chi}{z},\,
    \textstyle g=\frac{3\chi(1-\chi)}{\chi z^3+(1-\chi)(1+z)^3}\,.\label{eq:pre:long_fgStep}
  \end{split}
\end{equation}

In the large-gap regime while $S^{2/3}$ scaling is recovered for all $\chi$, the inlet length does control the magnitude high-$S$ limit. Using the profile, \Eqref{eq:pre:stepxi}, to evaluate $g(z)$, \Eqref{eq:pre:delta}, the thick film scaling limit $g \to 3\chi(1-\chi)/z^3$, with $f \to 1/z$ trivially. The resultant Stribeck curve is then,
\begin{equation}
    \mu(S) = \frac{d}{D} \times \frac{S^{2/3}}{[{3\chi(1-\chi)}]^{1/3}}\,.
\end{equation}
Friction is minimised in the large gap limit when the denominator is maximised, \ie\ $\chi = 0.5$, and $\mu$ diverges as $\chi \to 0$ or $1$~\cite{Note4}
Pertinent to our scaling analysis, the $\chi$-dependent factor is again of $\order{1}$ for $\chi \approx 0.01$~to~0.99.

As $z \rightarrow 0$, $g \to 3\chi$ for long bearings, \partFigref{fig:pre:longSteps}{e}, as the first two powers cancel, reducing the apparent $z^{-2}$ divergence, \Secref{sec:pre:powerExpand}. Meanwhile, the drag force [\Eqref{eq:pre:Flong}] $f \to (1-\chi)/z$ as only the second terms $\propto z^{-n}$ in \Eqref{eq:pre:stepxi} diverge. The small-$z$ limits are summarised in Table~\ref{tab:pre:limits}. As detailed in \Secref{sec:pre:calcConf}, while the equations remain analytic for the short bearing the lubrication approximation breaks down.




%


\end{document}